\documentclass{osa-article}

\usepackage{booktabs}
\usepackage{adjustbox}

\journal{boe}
\articletype{Research Article}
\begin{document}

\title{Corneal endothelium assessment in specular microscopy images with Fuchs' dystrophy via deep regression of signed distance maps}

\author{Juan S.~Sierra,\authormark{1} Jesus Pineda,\authormark{2} Daniela Rueda,\authormark{3} Alejandro Tello,\authormark{4,5,6,7} Angélica M.~Prada,\authormark{4,5} Virgilio Galvis,\authormark{4,5,6} Giovanni Volpe,\authormark{2} Maria S. Millan,\authormark{8} Lenny A. Romero,\authormark{9} and Andres G.~Marrugo\authormark{1,*}}

\address{\authormark{1}Facultad de Ingeniería, Universidad Tecnológica de Bolívar, Cartagena, Colombia\\
\authormark{2}Department of Physics, University of Gothenburg, SE-41296 Gothenburg, Sweden\\
\authormark{3}Hospital Oftalmológico Dr. Elías Santana en Santo Domingo,  República Dominicana\\
\authormark{4}Centro Oftalmológico Virgilio Galvis, Floridablanca, Colombia\\
\authormark{5}Fundación Oftalmológica de Santander FOSCAL, Floridablanca, Colombia\\
\authormark{6}Facultad de Salud, Universidad Autónoma de Bucaramanga UNAB, Bucaramanga, Colombia\\
\authormark{7}Facultad de Salud, Universidad Industrial de Santander UIS, Bucaramanga, Colombia\\
\authormark{8}Dept. Óptica y Optometría, Universidad Politécnica de Cataluña, Terrassa, Spain\\
\authormark{9}Facultad de Ciencias Básicas, Universidad Tecnológica de Bolívar, Cartagena, Colombia
}

\email{\authormark{*}agmarrugo@utb.edu.co} %% email address is required



\begin{abstract}

Specular microscopy assessment of the human corneal endothelium (CE) in Fuchs' dystrophy is challenging due to the presence of dark image regions called guttae. This paper proposes a UNet-based segmentation approach that requires minimal post-processing and achieves reliable CE morphometric assessment and guttae identification across all degrees of Fuchs' dystrophy. We cast the segmentation problem as a regression task of the cell and gutta signed distance maps instead of a pixel-level classification task as typically done with UNets. Compared to the conventional UNet classification approach, the distance-map regression approach converges faster in clinically relevant parameters. It also produces morphometric parameters that agree with the manually-segmented ground-truth data, namely the average cell density difference of -41.9 cells/mm$^2$ (95\% confidence interval (CI) [-306.2, 222.5]) and the average difference of mean cell area of 14.8~$\mu m^2$ (95\% CI [-41.9, 71.5]). These results suggest a promising alternative for CE assessment. 
\end{abstract}

%%%%%%%%%%%%%%%%%%%%%%%%%%  body  %%%%%%%%%%%%%%%%%%%%%%%%%%
\section{Introduction}

% The human corneal endothelium (CE) is responsible for maintaining corneal transparency and its proper hydration, both critical for good vision. It can be imaged in-vivo with specular microscopy and the resulting images can be analyzed to obtain clinical information by quantifying cell morphometric parameters. The most relevant are: cell density (CD), mean cell area (MCA), hexagonality (HEX\%), and coefficient of variation in cell area (CV\%). However, this quantification requires the accurate detection of cell contours, which is cumbersome and time-consuming to carry out manually.

The human corneal endothelium (CE) is responsible for maintaining corneal transparency and its proper hydration, both critical for good vision. It can be imaged \textit{in vivo} with specular microscopy, and the resulting images can be analyzed to obtain clinical information by quantifying cell morphometric parameters, like cell density~\cite{giasson2005morphometry}. However, this quantification requires the accurate detection of cell contours, which is especially challenging in the presence of corneal endotheliopathies, such as Fuchs’ dystrophy~\cite{giasson2005morphometry,sierra2020automated}. Commercially available software does not give satisfactory results~\cite{selig2015fully}. As an alternative, \textit{in vivo} confocal microscopy has been used to image the CE in Fuchs' dystrophy with remarkable results. However, it has not become a routine imaging device because it is contact-based and more technically challenging~\cite{Tone:2019}. Moreover, although anterior segment optical coherence tomography has been proposed for Fuchs' dystrophy grading~\cite{yasukura2021new}, it does not provide CE morphometric parameters, which specular microscopy does.

% The optical principle of specular microscopy is based on epi-illumination~\cite{laing1979clinical}. As shown in Fig.~\ref{fig:optical}(a), a slit of light is projected through the corneal tissue. Most of this light is transmitted into the aqueous humor (Fig.~\ref{fig:optical}b), some is reflected by the anterior and posterior corneal surfaces, and the rest of the light is scattered by the various corneal layers (Fig.~\ref{fig:optical}(c)). Due to the lack of luminosity and the rough surface of the CE, the portion of light reflected by the CE layer is diffuse, thus the objective lens can only collect a small amount of diffuse reflected light, producing an image of the CE cells like the one shown in Fig.~\ref{fig:optical}(d). Due to the eye curvature, the brightness of the formed image is not uniform, which makes these images even more difficult to be digitally processed.

\begin{figure}[t]
 \centering
 \includegraphics[width=0.6\textwidth]{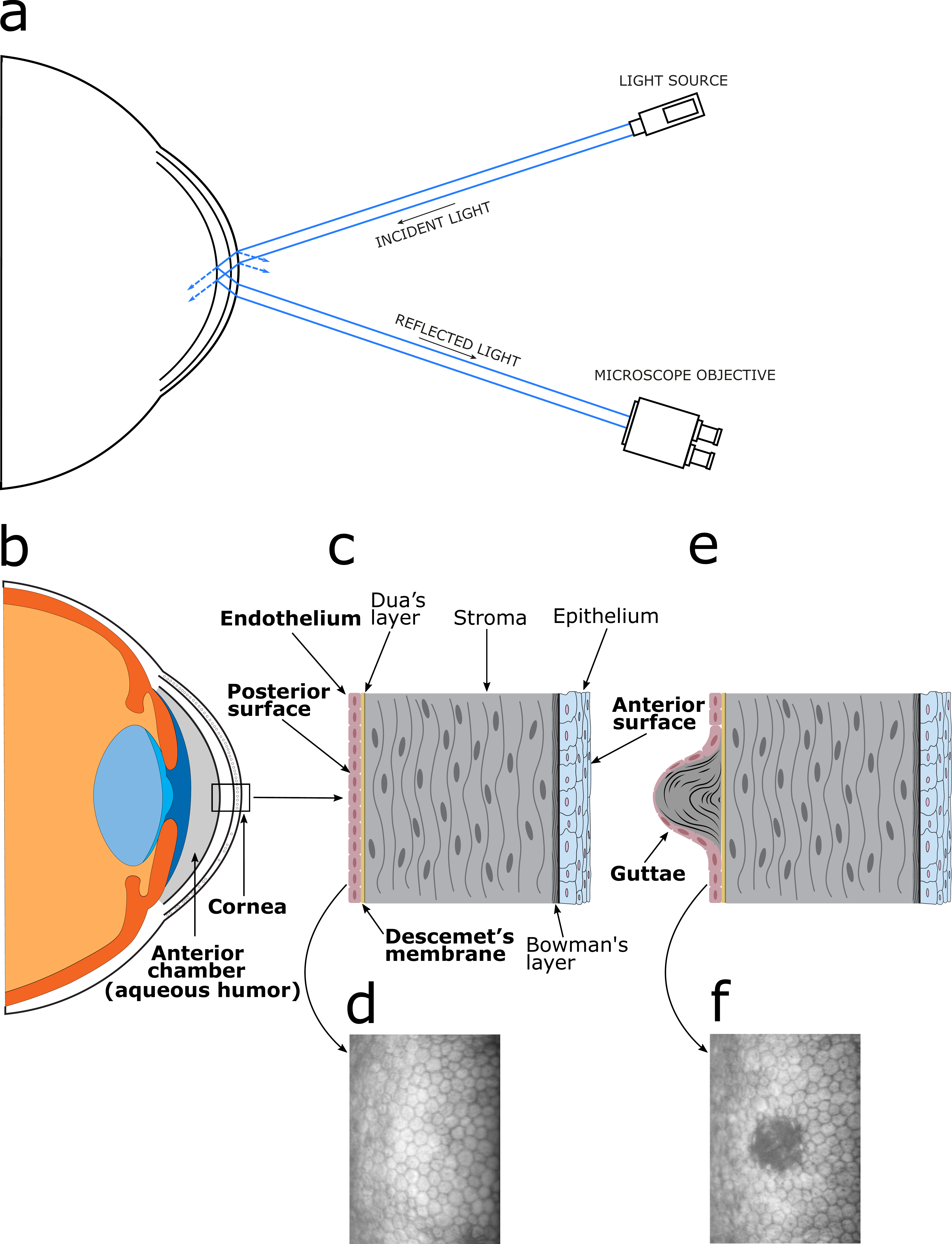}
 \caption{Specular microscopy imaging of the corneal endothelium (CE). \textbf{(a)} Optical principle. A light source projects light onto the corneal surface from which only a fraction is reflected by the endothelium and collected by the microscope. 
%  The remaining are either scattered or reflected by other interfaces. 
\textbf{(b)} The anterior segment including the cornea and the anterior chamber. \textbf{(c)} The corneal layers including the endothelium. 
%  The epithelium, Bowman's layer, stroma, Dua's layer, Descemet's membrane, and the endothelium.
\textbf{(d)} Typical CE image. The cells are uniformly distributed along the entire CE tessellation. \textbf{(e)} Example of a corneal gutta showing the outgrowth produced due to abnormal CE cells. \textbf{(f)} CE image with Fuchs' dystrophy with a large gutta in the center.}
 \label{fig:optical}
\end{figure}

Specular microscopy is based on illuminating the cornea with a narrow beam of light (Fig.~\ref{fig:optical}(a)) and capturing the specular reflection from the posterior corneal surface~\cite{laing1979clinical}. Most of the incident light is either transmitted into the eye's anterior chamber (Fig.~\ref{fig:optical}(b)) or reflected by the epithelial surface at the anterior surface of the cornea. None of these fractions of the incident light are useful for acquiring the endothelial image but the tiny fraction (about 0.22\%~\cite{Srinivasan_2005}) reflected by the posterior corneal surface (Fig.~\ref{fig:optical}(c)). The CE image is a trade-off between the width of the beam and the corneal thickness. Once acquired, the normal endothelial cells appear gray, forming a regular tessellation (Fig.~\ref{fig:optical}(d)). The irregularities in the surface (Fig.~\ref{fig:optical}(e)) produce reflected rays in directions other than the corresponding specular reflection and, consequently, appear as dark regions (Fig.~\ref{fig:optical}(f)).

Computer vision techniques are often used to carry out the CE cell segmentation task~\cite{nurzynska2018deep}. However, developing a fully automated method to assess CE health is currently a challenge in ophthalmology~\cite{fabijanska2018segmentation}.
Scarpa and Ruggeri~\cite{scarpa2016automated,scarpa2016development} proposed an automated method for cell segmentation using a genetic algorithm that combines information about the typical regularity of endothelial cell shape with the intensity of the image pixels. Watershed algorithms and morphological operations applied on thresholded images have been frequently used to perform cell segmentation~\cite{sanchez1999automatic,selig2015fully,al2018fully,piorkowski2017influence}. In contrast, other methods estimate cell density without needing segmentation by using spatial frequency analysis~\cite{selig2015fully} or two-dimensional discrete Fourier transforms~\cite{ruggeri2005new}. However, these methods are not immediately applicable in CE images with Fuchs' dystrophy. At the same time, other authors have studied CE morphometry in both normal and dystrophic corneas. For instance, Giasson \textit{et al.},~\cite{giasson2005morphometry} developed a contour detection algorithm based on morphological image transformations to quantify cells and guttae. However, the method required significant manual interaction.

% Deep learning algorithms have reached high popularity in a wide range of applications. These algorithms can learn patterns from a given set of training samples. Supervised learning methods require labeled datasets for learning the mapping from the input image to the desired output. More precisely, convolutional neural networks (CNNs) are one of the best models because of their capacity for feature extraction~\cite{vicar2021self}. Medical imaging, including ophthalmology, is one field that has benefited from using these algorithms, improving the results of standard methods in terms of accuracy and robustness. M. C. Daniel \textit{et al.}~\cite{daniel2019automated} assessed the performance of the UNet, a CNN architecture for biomedical image segmentation~\cite{ronneberger2015u}. They used a large dataset containing CE images of different quality and clinical conditions; however, their analysis focused on cell characterization despite the presence of guttae. Other authors have followed a similar approach for dealing with the cell segmentation problem~\cite{fabijanska2018segmentation,joseph2020quantitative}; however, to the best of our knowledge, none have characterized the guttae simultaneously.

Recent methods based on deep learning have achieved considerable improvements. More precisely, deep convolutional neural networks (CNN) are used for their capacity for feature extraction~\cite{vicar2021self}. Daniel \textit{et al.}~\cite{daniel2019automated} assessed the performance of the UNet (a CNN architecture for biomedical image segmentation \cite{ronneberger2015u}) in CE segmentation. They used a large dataset containing CE images of different quality and clinical conditions. However, their analysis focused on cell characterization despite the presence of guttae. Other authors have followed a similar approach to dealing with the cell segmentation problem~\cite{nurzynska2018deep,fabijanska2018segmentation,joseph2020quantitative}. Nevertheless, simultaneous automated characterization of CE cells and guttae remains a difficult problem, and guttae parametrization provides an opportunity for improving CE assessment.

% Other works include using an artificial neural network to classify pixels between cell body and cell contour applied on CE images acquired from a bank of corneas of porcine eyes stained with alizarine red~\cite{ruggeri2010system}. A similar approach was proposed by K. Nurzynska~\cite{nurzynska2018deep}, where the author used a CNN to classify pixels between cell center, cell body and cell border to achieve precise cell segmentation. Vigueras-Guillén \textit{et al.}~\cite{vigueras2019fully,vigueras2020deep} used a fully convolutional architecture based on the UNet model and a sliding-window CNN to assess the CE to detect cells edges. They also implemented a densely connected UNet architecture to find the region of interest (ROI) on specular microscopy images, where individual cells are easily recognizable\cite{vigueras2019automatic}, and developed a CNN-based regression to estimate biomarkers in specular microscopy images, which combines the previously mentioned methods~\cite{vigueras2019convolutional}. Vigueras-Guillén \textit{et al}.~\cite{vigueras2022denseunets} also proposed an attention mechanism called feedback non-local attention to infer cell edges in CE images with guttae with accurate results; however, it required a large amount of manually segmented CE images, the use of more than one deep learning model, a complex post-processing and heuristics.

Other methods based on neural networks have explored different codification strategies to improve performance. Vigueras-Guillén \textit{et al.}~\cite{vigueras2019fully,vigueras2020deep} used a fully convolutional architecture based on the UNet model and a sliding-window CNN to assess the CE image for detecting cell edges. They also implemented a densely connected UNet architecture to find the region of interest (ROI) on specular microscopy images, where individual cells are easily recognizable\cite{vigueras2019automatic}. They developed a CNN-based regression to estimate biomarkers in specular microscopy images, which combines the previously mentioned methods~\cite{vigueras2019convolutional}. Vigueras-Guillén \textit{et al}.~\cite{vigueras2022denseunets} also proposed an attention mechanism called feedback non-local attention to infer cell edges in CE images with guttae for improving accuracy results. However, this method required many manually segmented images, using more than one deep learning model and complex post-processing and heuristics.

% Nevertheless, existing automated software often fails due to severe corneal endothelial dysfunction, such as Fuchs’ endothelial corneal dystrophy, which is one of the most common corneal diseases\cite{sierra2020automated}. The global prevalence rate of Fuchs’ Dystrophy has been estimated to be 7.33\%~\cite{aiello2022global}. However, in populations above 50 years old the rate increases to 9.20\%, being 2.2-times more likely in women than in men. Fuchs' dystrophy is related to the accumulation of collagen secreted by abnormal endothelial cells to the Descemet's membrane, as shown in Fig.~\ref{fig:optical}(c). It produces outgrowths that protrude to the anterior chamber, also called guttae~\cite{feizi2018corneal,eghrari2015fuchs,laing1981endothelial}. In specular microscopy, they appear as dark regions without identifiable cells, as shown in Fig.~\ref{fig:correction}(a), where cells are altered and apparently non functional~\cite{chiou1999confocal,hogan1974fuchs,iwamoto1971electron,tone2021fuchs}. These guttae are produced due to the difference, in terms of depth, between the reflection plane and the protrusion (it is out of focus), and they often hinder automatic segmentation.

\begin{figure}[t]
 \centering
 \includegraphics[width=0.65\linewidth]{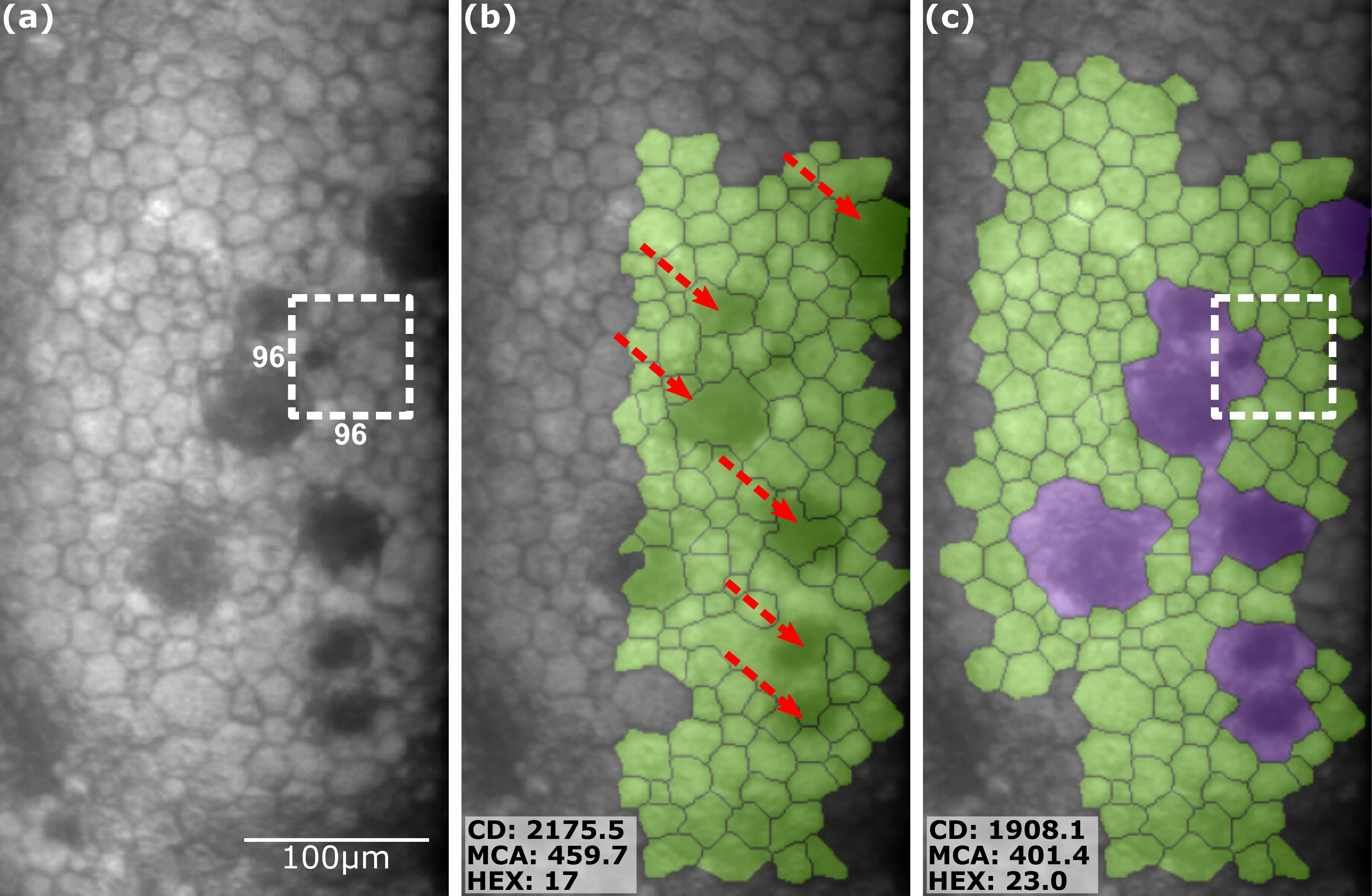}
 \caption{\textbf{(a)} A CE image with Fuchs' dystrophy.  \textbf{(b)} The segmentation performed by the specular microscope software. The green color indicates regions labeled as cells. The software misclassifies abnormal regions (guttae) as large cells, as shown with red arrows. \textbf{(c)} The manually annotated ground-truth reference. The violet color represents the guttae. The microscope image and the segmentation were split into $96\times96$ pixels image patches to generate the training and validation data sets. The estimated parameters are shown at the bottom left corner, indicating a significant difference.} 
 \label{fig:correction}
\end{figure}

Nonetheless, existing automated software often fails due to severe corneal endothelial dysfunction, such as Fuchs’ dystrophy, one of the most common corneal diseases\cite{sierra2020automated}. The global prevalence rate of Fuchs’ dystrophy is around 7\%~\cite{aiello2022global}. However, in populations above 50, the rate increases to 9\%, being 2.2-times more likely in women than in men. Fuchs' dystrophy is related to the accumulation of collagen secreted by abnormal endothelial cells to the Descemet's membrane, as shown in Fig.~\ref{fig:optical}(c). It produces outgrowths that protrude to the anterior chamber, also called guttae~\cite{feizi2018corneal,eghrari2015fuchs,laing1981endothelial}. In specular microscopy, they appear as dark regions without identifiable cells, as shown in Fig.~\ref{fig:correction}(a), where cells are altered and apparently non functional~\cite{chiou1999confocal,hogan1974fuchs,iwamoto1971electron,tone2021fuchs}. These guttae are produced due to the depth difference between the reflection plane and the protrusion~(Fig.~\ref{fig:optical}(e)). Accurately characterizing the CE stage with new parameters that include guttae should make progression follow-up more precise. 

% Descemet's membrane is one of the six layers that compose the cornea and is located between the stroma and the corneal endothelium. It contains different kinds of collagen secreted by the endothelial cells. Often, abnormal cells produce an accumulation of collagen that appear as outgrowths on the posterior surface of Descemet's membrane, as shown in Fig.~\ref{fig:optical}(e). It is the main characteristic of the endothelial tessellation in Fuchs'  dystrophy, also called  \textit{cornea guttata}~\cite{laing1981endothelial}. Dark regions in cornea guttata images (like those that we can see in Fig.~\ref{fig:optical}(f)) are produced due to the difference, in terms of depth, between the reflection plane and the protrusion (it is out of focus). It may lead to inaccurate automatic segmentations.

In this work, we propose a deep learning-based method to carry out reliably the segmentation task in specular microscopy images in the presence of cornea guttata. We use a CNN architecture based on the UNet model for mapping the input image to a signed distance map, from which we obtain the cell and guttae segmentation. Our network demonstrates rapid convergence and robustness in terms of clinically relevant CE morphometric measures. We evaluate the main CE morphometric parameters necessary to estimate its health status and compare them with manual references. Moreover, we compare the results with the evaluation performed by the \textit{CellCount} microscope software from Topcon (where the cell size-dependent parameters are usually overestimated, as shown in Fig.~\ref{fig:correction}). Our results show an improvement over conventional UNet-based methods and the Topcon software used routinely in the clinical setting. 

%%%%%%%%%%%%%%%%%%%%%%%%%%%%%%%%%%%%%%%%%%%%%%%%%%%%%
\section{Materials and Methods}

\begin{figure}[!ht]
 \centering

 \includegraphics[width=0.85\linewidth]{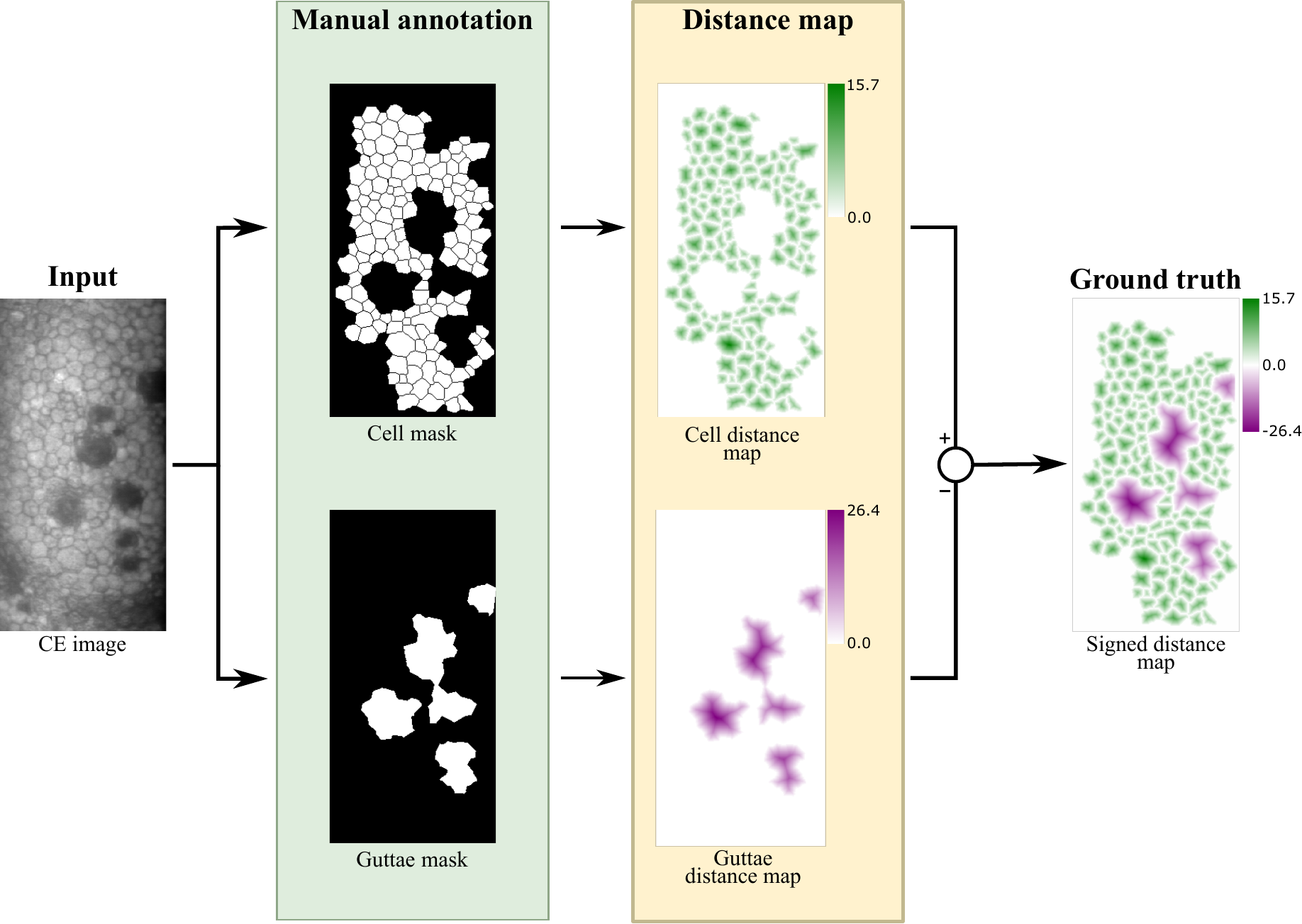}
 \caption{Data annotation stage. A trained physician manually annotates the segmentation using our custom-built software to produce the masks for cells and guttae. Then, a distance transform is applied on the two masks. Finally, negative values are assigned to the guttae distance map resulting image in a signed distance map,  with positive values for cells (green) and negative values for guttae (violet).}
 \label{fig:method}
\end{figure}

We cast the problem of corneal endothelium (CE) health evaluation as a supervised regression~\cite{he2021deeply}. We used a CNN based on the UNet architecture to predict a signed distance map from a given CE image. A high-level description of
the proposed codification strategy is shown in Fig.~\ref{fig:method}. The input CE images were acquired with a specular endothelial microscope (SP-3000P, Topcon Co., Japan; magnification 150$\times$, and image size of 0.25 $\times$ 0.5 mm). Each image was processed by the Topcon \textit{CellCount} microscope software to perform an initial segmentation, which can be used to create ground truths. This initial segmentation had errors in various corneal regions, particularly in corneas with Fuchs' dystrophy. Therefore, the initial segmentations were manually curated using the \textit{Data Annotation} custom built software, as shown in Fig.~\ref{fig:gui}. A distance transform was applied to the curated segmentations to generate the signed distance maps with which we train the network. We used thresholding and watershed transform to post-process the model output and calculate the main morphometric parameters. 
% Finally, the network was trained and validated with the curated data. 
A detailed description of the proposed method is explained below.
The study protocol was aproved by the ethics committee of the Universidad Tecnológica de Bolívar, Colombia, and the requirement for informed consent was waived because of the retrospective study design. The study adhered to the tenets of the Declaration of Helsinki.

% The Institutional Review Board of the  approved this study, and the requirement for informed consent was waived because of the retrospective study design. The study adhered to the tenets of the Declaration of Helsinki.

%It has a physical meaning

\subsection{Data Collection}

We used a set of 90 in vivo specular microscopy images of CE acquired from 66 patients with both healthy (42) and dystrophic corneas (48), using a Topcon SP-3000P specular microscope. The set of images was divided into three data-sets: training (57), validation (10) and testing (23). The procedure was realized in automatic mode with the \textit{CellCount} software equipped in the microscope, which provides an initial segmentation of the CE in a selected ROI. This software misclassifies guttae as cells, as shown in Fig.~\ref{fig:correction}(b). It can be modified with editing tools in the microscope software, for instance, to draw or remove cells. However, removing erroneous detections may lead to over- or under-estimation of morphometric cell parameters. The microscope exports a two-channel TIFF file of $640 \times 480$ pixels that contains the acquired CE image and the corresponding initial segmentation.

\subsection{Data Annotation}
\label{sec:seg_tool}

\begin{figure}[t]
 \centering
 \includegraphics[width=0.7\linewidth]{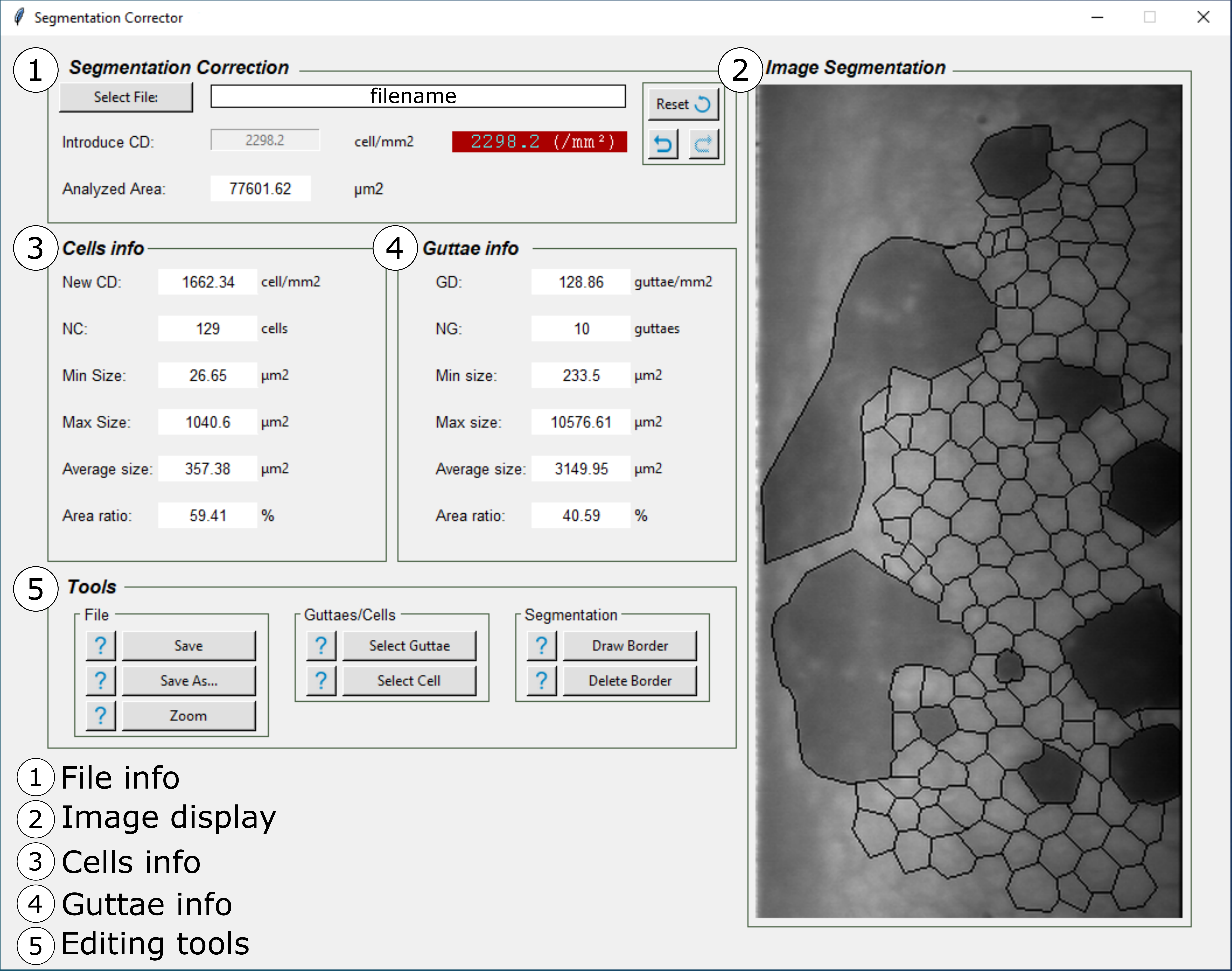}
 \caption{GUI of the data annotation software developed for this research. Panel~1 shows basic information about the loaded file. Panel 2 shows the CE image overlayed with the current segmentation. Panels 3 and 4 display the CE calculated parameters, and panel 5 contains the editing tools.}
 \label{fig:gui}
\end{figure}

Since the proposal of this work is a supervised deep learning-based approach, a curated dataset of CE segmentations is needed for training. Therefore, we developed a custom-built software to manually segments cells in CE images~\cite{sierra2020generating} to generate ground-truths. The software was made using the Python-based Tkinter library~\cite{lundh1999introduction} to create the graphical user interface (GUI) shown in Fig.~\ref{fig:gui}.

The image on the GUI's right side is composed of the current segmentation overlaid on top of the CE image. The software allows a trained ophthalmologist to modify the initial segmentation made by the \textit{CellCount} Topcon software, i.e., split or merge regions and classify these regions as cells or guttae. However, this segmentation is imported only to assist the ophthalmologist and reduce the time and effort spent on the annotation task. It can also be discarded to start the annotation from scratch. These tools are based on morphological operations applied to the binary image that contains the segmentation. After finishing the corrections, the software produces two masks corresponding to cell bodies and guttae/dystrophic regions, respectively, as shown in Fig.~\ref{fig:method}. From these images, all parameters are calculated: cell/guttae density, number of regions, minimum, maximum, and average area, and percentage of area occupied by each class in the segmented region. Finally, the corrected segmentation is saved as a three-page TIFF file with the new corrected segmentation. Six physicians defined the segmentation criteria, and three worked on independent subsets of 30 CE images to create the ground truths.

\subsection{Distance Maps}

The main problem in microscopy image cell segmentation is that close or overlapping cells tend to be segmented as a single object. Moreover, since dysfunctional regions or guttae often produce false positives in conventional cell segmentation software, we need to establish a reliable approach to discern between the two features while separating individual cells effectively. Therefore, we cast the supervised learning problem as a regression of an input image to a signed distance map. 

Typically, UNet-based models are trained under a supervised classification framework in which the network has to directly output a mask of different labels corresponding to the classes of objects in the image. This mask is computed via the softmax output of the neural network. However, often many stages of post-processing with heuristics need to be carried out to obtain the desired segmentation~\cite{vigueras2022denseunets, vigueras2019fully}. To avoid these post-processing stages we propose training our UNet with signed distance maps, which have several advantages, mainly: i) accurate segmentation of touching objects like cells; ii) robustness to unreliable ground-truth segmentations which are often problematic due to  poorly-defined cell or guttae boundaries; iii) continuity and smoothness constraints implied in estimating distance maps facilitate the detection of guttae, especially when they cover a significant area of the image~\cite{grauer2021active}; iv) 
reduced learning complexity and rapid convergence using a relatively small dataset. The network is intended to predict Euclidean distances between the center and edge of cells/guttae rather than a pixel-wise classification of the input image, which is prone to incorrect classification due to the the lack of uniform illumination, poor contrast, and cell visibility, and requires a large number of images to develop reliable mapping capabilities \cite{vigueras2022denseunets}. 
% To show the advantages of the distance map approach, we trained the UNet under these two scenarios, which are explained in the next section.

The signed distance maps are created as follows: given a grayscale CE image as input, the goal is to train a CNN model to predict a signed distance map where positive values indicate cell bodies and negative values guttae. For preparing the target signed distance maps, the reference segmentations (cell and guttae masks) are passed through a distance transform that assigns to each pixel the value of the Euclidean distance to the closest background pixel~\cite{naylor2018segmentation}. This procedure produces two distance maps, as depicted in Fig.~\ref{fig:method} where the green map corresponds to cells and the violet map to guttae. The final signed distance map $\mathcal{D}_I$ is given by
\begin{align}
    \mathcal{D}_I = \mathcal{D}_c - \mathcal{D}_g,
\end{align}
where $\mathcal{D}_c$ is the calculated distance map from the cell mask, and $\mathcal{D}_g$ is the distance map corresponding to guttae regions. It encodes cells as positive values and guttae as negative values in a single scalar field image. The resulting signed distance map will feature larger values for larger regions~\cite{naylor2018segmentation}, which, for this work, may be useful due to the significant differences between region sizes.

\begin{figure}[t]
\centering
 \includegraphics[width=\linewidth]{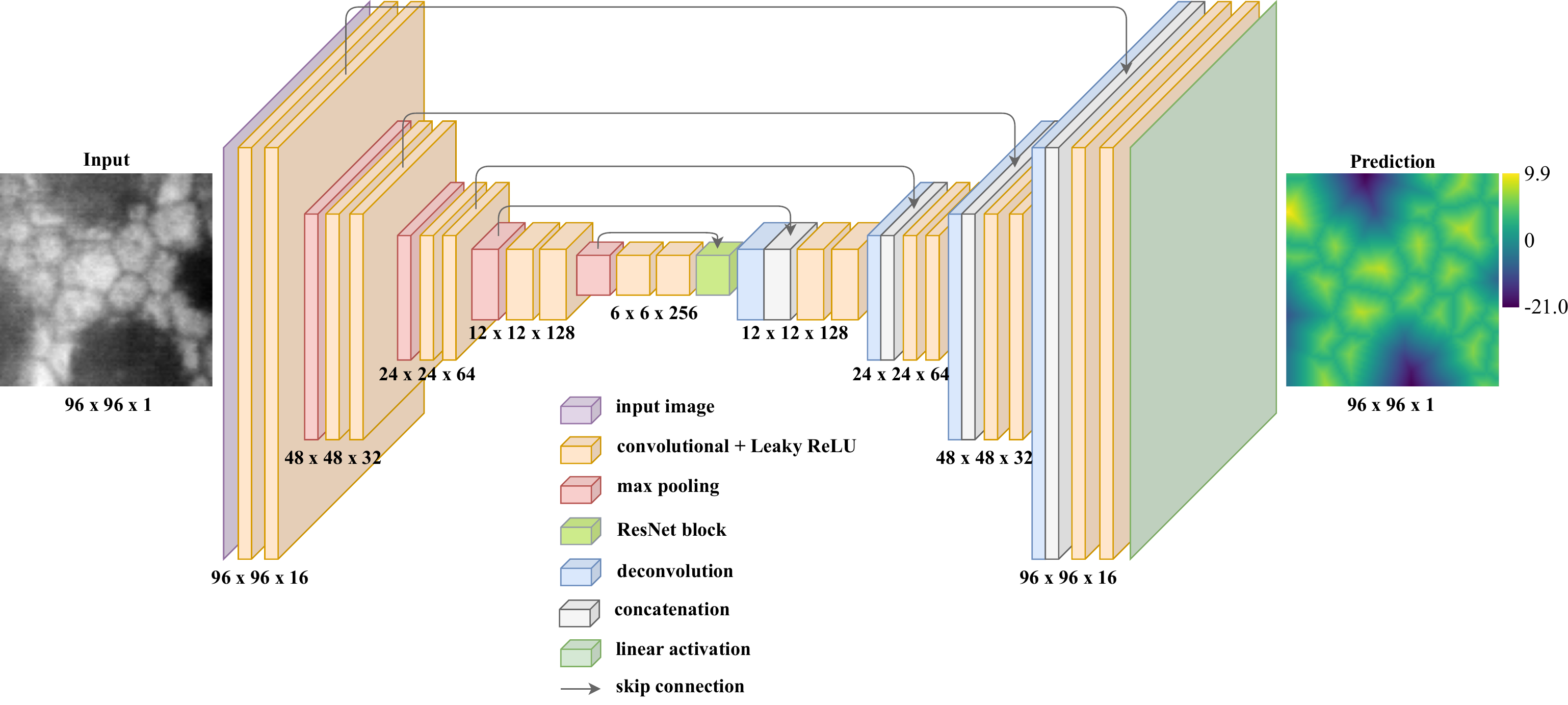}
 \caption{\textbf{Schematic of the CNN architecture}. Each encoding path layer is made up of two sequences of $3 \times 3$ convolution blocks. Each block consists of Instance Normalization and Leaky ReLU activation with negative slope coefficient $\alpha = 0.1$. It is followed by a $2 \times 2$ max-pooling layer with a stride of 2 for downsampling. The decoding path consists of deconvolution layers (kernel size of $2 \times 2$ and stride of $2$) followed by two convolution blocks. Skip connections allow transferring data between layers of the same level in the downsampling/upsampling paths. The last layer is computed by a $1 \times 1$ convolution with linear activation to perform the output distance map. We trained the CNN using the MAE loss function and Adam optimizer.}
 \label{fig:model}
\end{figure}  

\subsection{CNN Architecture and Implementation}

To develop a robust algorithm for the CE image segmentation task, we aimed to use a deep learning-based method that would deliver accurate cell parameters and guttae information. For this purpose, we devised a a 5-layers UNet architecture to predict signed distance maps. As shown in Fig.~\ref{fig:model}, the model consists of two stages: encoding and decoding. The encoding stage (left side) follows the standard CNN: it consists of convolutional blocks, each followed by a $2 \times 2$ max-pooling operation with a stride of 2 for downsampling. Each convolutional block contains two sequences of $3 \times 3$ convolution layers with padding, Instance Normalization~\cite{ulyanov2016instance}, and leaky rectified linear unit (Leaky ReLU) activation~\cite{maas2013rectifier,xu2015empirical} with negative slope coefficient $\alpha = 0.1$. At each downsampling step the number of feature channels is doubled. The downmost layers are ResNet blocks, i.e., the input of the block is added to the output~\cite{he2016deep}. The use of these blocks is motivated by the need to tackle the vanishing gradient problem and to improve the latent space representation~\cite{helgadottir2021extracting}. 

The decoding stage (the expansive path in the right side) is similar, except that max-pooling operations are replaced by transposed convolution (deconvolutions) layers with a kernel size of $2 \times 2$ and a stride of 2, and the number of feature channels is halved. Each deconvolution is followed by concatenation with the corresponding feature map from the encoding path to recover spatial information lost due to the downsampling operations, and a convolution block as in the encoding stage. At the final layer, a $1 \times 1$ convolution with linear activation is used to map each 16-component feature vector to the desired output. This network was implemented using the Python-based Keras library with a TensorFlow backend, and the Python-based DeepTrack library~\cite{midtvedt2021quantitative,helgadottir2019digital}.  

Before training, each patch is z-scored and normalized between -1 and 1 with the $\tanh(.)$ function. Then, during the training process, simple data augmentation operations are used, such as horizontal, vertical, and diagonal flips, and random rotations, to increase the number of samples from 148 to 2048, combined with the mean absolute error (MAE) loss function, and Adam optimizer with a learning rate of 0.001~\cite{kingma2014adam}. Also, a continuous generator continuously creates new images during training by balancing the speed gained from reusing images and the generalization achieved from yielding new training data. We trained the model on a Tesla T4 GPU with 12GB of RAM. Training time takes, on average, 33 minutes. The model code is available in a GitHub repository specified in the code availability statement.

% %
% \begin{equation}
%     \label{eq:loss}
%     \centering
%     loss(\mathcal{D}_{\mathcal{I}}, \hat{\mathcal{D}}_{\mathcal{I}}) = \frac{1}{\text{number of pixels}} \sum_{x,y} |\mathcal{D}_{\mathcal{I}}[x,y] - \hat{\mathcal{D}}_{\mathcal{I}}[x,y]|
% \end{equation}
% % 

\subsection{Post-processing stage}

The output distance map predicted by the CNN is then binarized to separate guttae from cells as
% 
% with fixed empirically-found thresholds to obtain
% 
\begin{equation}
    \centering
    \mathcal{T}_g(i,j) = 
        \begin{cases}
          1, & \text{if}\ \mathcal{D}_{\mathcal{I}}(i,j) < 0 \\
          0, & \text{otherwise}.
        \end{cases}
        \label{eq:guttae}
\end{equation}
However, if we use the condition $\mathcal{D}_{\mathcal{I}}(i,j) > 0$ we may get some cells touching each other. We found that a threshold slightly higher than 0 for cells would separate each cell from the surrounding ones for easier counting and post-processing. Therefore, we set this threshold empirically to 0.2, as
\begin{equation}
    \centering
    \mathcal{T}_c(i,j) = 
        \begin{cases}
          1, & \text{if}\ \mathcal{D}_{\mathcal{I}}(i,j) > 0.2 \\
          0, & \text{otherwise}
        \end{cases}
    \enspace
\end{equation}
to identify cells.
Finally, we calculate $\mathcal{T}(\mathcal{D}) = \mathcal{T}_c \cup \mathcal{T}_g$ to perform the watershed transformation to ensure that boundaries between cells and guttae are well defined. The result after this process is the specular microscopy segmentation, which is separated into cells and guttae regions, avoiding complex post-processing operations.

\subsection{Morphometric Parameters}

\begin{table}[ht]
\centering
\caption{Summary of the main morphometric parameters used to assess the corneal endothelium health.}
\label{tab:parameters}
\begin{tabular}{@{}p{3cm}p{2.8cm}p{6.7cm}@{}}
\toprule
Parameter &
  Definition &
  Clinical relevancy \\ \midrule
Cell density (CD). &
  The number of cells per unit area, measured in cells/mm$^2$.  &
 CD < 1000 cells/mm$^2$ is considered a risk factor for corneal edema, and CD < 500 cells/mm$^2$ frequently leads to corneal decompensation and clinically significant stromal edema. Typical CD in healthy adults over 60 is $\sim 2000$ cells/mm$^2$ \cite{roszkowska2004age,valdez2022age}. \\
Mean cell area (MCA). &
  Average cell size measured in µm$^2$. &
  The average cell size in adults over 60 is usually between 400 and 450 µm$^2$ \cite{valdez2022age}. Larger cells are correlated with lower cellular densities. \\
Hexagonality (HEX\%), also called pleomorphism. &
  Percentage of hexagonal cells. &
A healthy endothelium in an adult over 60 will have roughly 45\% of hexagonal cells \cite{valdez2022age}. Endothelial cell stress will cause cell loss, and some of the remaining cells will increase their area and change their shape leading to a HEX\% change.
\\
Coefficient of variation of cell area (CV\%), also called polymegethism &
  Standard deviation of cell area divided by the average cell area, to be reported as a percentage. &
  A healthy endothelium in an adult over 60 will have roughly 45\% of CV\% \cite{valdez2022age}. Endothelial cell stress will cause cell loss, and some of the remaining cells will increase their area, but some will remain small, leading to a CV\% change \cite{kudva2020corneal}.
  \\
Guttae Area Ratio (GAR\%) proposed parameter. &
  Total guttae area as a ratio of total segmented area. &
  This new parameter indicates the proportion of analyzed areas covered by guttae (areas devoid of cells or covered by altered cells). Thus, the percentage of the analyzed area they cover is an indicator of the degree of endothelium alteration~\cite{shilpashree2021automated}.\\
  \bottomrule
\end{tabular}
\end{table}

Here, we describe the most relevant morphometric parameters used to assess the CE state in the clinical setting and use them to evaluate the method's performance~\cite{ahmed2022comparison}. Cells touching the borders have been automatically removed for calculating the parameters because they may correspond to a partial segmentation. 
% Most literature on cell segmentation  quantitative assessment-based classification error or intersection-over-union type metrics, however, these are often prone to problems like individual cell identification or inaccurate boundary estimation~\cite{sierra2020automated,al2018fully,ahmed2022comparison}. 
% We avoided this problem by using the following morphometric parameters shown in Table~\ref{tab:parameters} to assess the agreement between methods. 
The cell density (CD), given by
\begin{equation}
    \label{eq:cd}
    \centering
    \text{CD} = \frac{\#~\text{of cells}}{\text{total segmented area}} \enspace,
\end{equation}
which indicates the number of cells per unit area, measured in cells/mm$^2$. It is noteworthy that this parameter is often incorrectly estimated in the presence of cornea guttata, because the effective area has to include the area occupied by guttae or dystrophic regions, but the numerator has to account for healthy cells, while discounting any guttae misclassified as cells. The mean cell area~(MCA), given by
\begin{equation}
    \label{eq:mca}
    \centering
    \text{MCA} = \frac{\text{total cell area}}{\#~\text{of cells}} \enspace,
\end{equation}
measured in $\mu \mathrm{m}^2$.
This parameter is also prone to over-estimation in cornea guttata. Hexagonality (HEX\%) also called pleomorphism, i.e., the percentage of hexagonal cells (cells with six neighbors), calculated as
\begin{equation}
    \label{eq:hex}
    \centering
    \text{HEX}\% = \frac{\#~\text{of hexagonal cells}}{\#~\text{of cells}} \times 100\% \enspace.
\end{equation}
The coefficient of variation of cell area (CV\%), also called polymegethism, calculated as
\begin{equation}
    \label{eq:cv}
    \centering
    \text{CV}\% = \frac{\text{std}(\text{cells area})}{\text{mean}(\text{cells area})} \times 100\% \enspace,
\end{equation}
where std(.) and mean(.) are the standard deviation and the mean of the cell areas, respectively. 

Finally, we report a new parameter to quantify the percentage of the segmented area affected by guttae. We called it the Guttae Area Ratio (GAR\%), calculated as
\begin{equation}
    \label{eq:gar}
    \centering
    \text{GAR}\% = \frac{\text{total guttae area}}{\text{total segmented area}} \times 100\% \enspace.
\end{equation}
We believe this parameter provides a complementary CE assessment tool for the clinician~\cite{shilpashree2021automated}.
In Table~\ref{tab:parameters}, we briefly describe the clinical relevancy of these parameters and how they are typically affected in CE images with Fuchs' dystrophy.

%%%%%%%%%%%%%%%%%%%%%%%%%%%%%%%%%%%%%%%%%%%%%%%%%%%%%
\section{Results}

We used a set of 23 images to evaluate the performance of four methods: the Topcon microscope \textit{CellCount} software, the custom-built software (reference segmentation), the CNN-architecture shown in Fig.~\ref{fig:model} trained using masks (UNet-mask)~\cite{shilpashree2021automated} and the proposed method trained using signed distance maps (UNet-dm). First, we show a comparison between the results obtained by the UNet-mask and the UNet-dm. Second, we discuss a qualitative comparison between several representative images of four grades of endothelial Fuchs dystrophy (grade 0 for a cornea without guttae, grade 1 for mild grade cornea guttata, grade 2 for moderate level, and grade 3 for severe cases of cornea guttata). The agreement was determined using the morphometric parameters described above and Bland-Altman plots.

\subsection{Classification versus regression UNets}

\begin{figure}[t]
\centering
 \includegraphics[width=\linewidth]{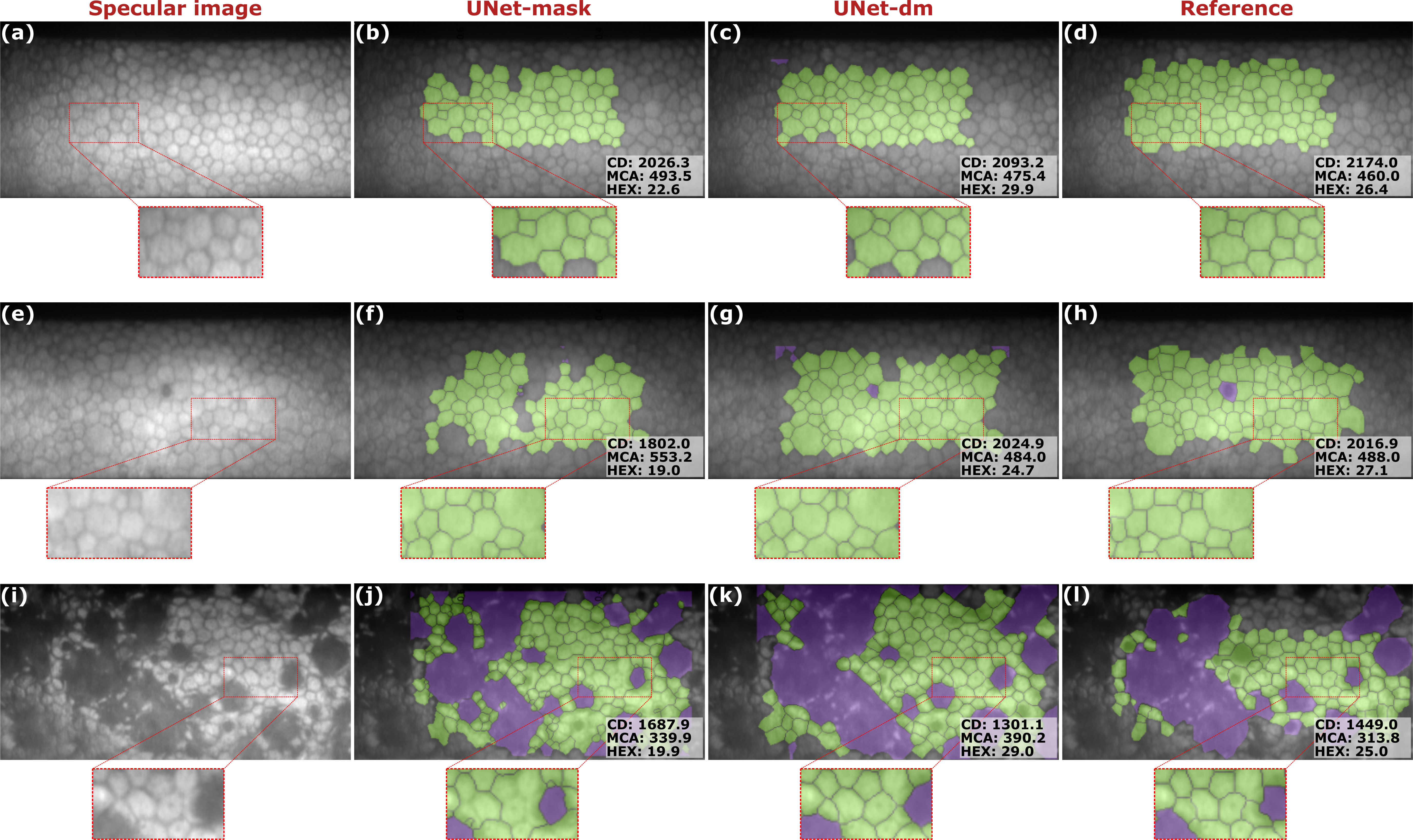}
 \caption{Comparison between UNet-mask and UNet-dm in one example of \textbf{(a-d)} non-guttae cornea, \textbf{(e-h)} grade 1 and \textbf{(i-l)} grade 3 of cornea guttata. Each result is composed of the specular microscopy image, the final segmentations with the UNet-mask and UNet-dm, and the reference segmentation (ground-truth). The estimated parameters are shown at the bottom right corner.}
 \label{fig:comparison}
\end{figure}

To show the advantages of the distance map approach, we trained the UNet under two scenarios: a classification (UNet-mask similar to~\cite{shilpashree2021automated}) and a regression approach (UNet-dm). Since the UNet-mask is a multi-class classification architecture, we had to make several modifications to the proposed method (Fig.~\ref{fig:model}): on the convolution layers, Instance Normalization was removed, and ReLU replaced Leaky ReLU activation. We also changed the loss function from MAE to weighted categorical cross-entropy. Finally, a softmax activation in the last layer computes the probability distribution over the three labels, i.e., cells, guttae, and intercellular space. The segmented image was post-processed with the watershed algorithm to separate cell boundaries.

The average accuracy was UNet-mask=83.06\% vs UNet-dm=83.79\%, which shows that both networks are able to identify sufficiently well the cells and guttae~\cite{yeghiazaryan2018family}. In Fig.~\ref{fig:comparison}, we show three examples of results obtained with UNet-mask and UNet-dm. Overall, both segmentations are similar in terms of the segmented areas. However, the UNet-mask results show problematic segmentations like merged cells. While there are ways to deal with these problems in post-processing, directly avoiding them through a distance-map codification is a substantial improvement. The results from the UNet-dm are much more similar to the ground-truth segmentations with well-defined cell boundaries. 

Moreover, the morphometric parameters obtained from the UNet-dm are closer to the reference values than those obtained with the UNet-mask. For instance, in the second and third row of Fig.~\ref{fig:comparison} the CD and HEX are underestimated.
% the estimated parameters are similar, in the first and third rows, the UNet-mask results show that the CD is considerably overestimated, while the MCA is underestimated. This issue is given by inadequate classification of pixels in a complicated area (lower contrast and illumination), like in the bottom-left corner of the segmentation shown in Fig.~\ref{fig:comparison}(j). Here, small regions of a few pixels are counted as individual cells. It produces the CD to increase, the MCA to decrease and affects the HEX\%.

\begin{figure}[t]
\centering
 \includegraphics[width=0.8\linewidth]{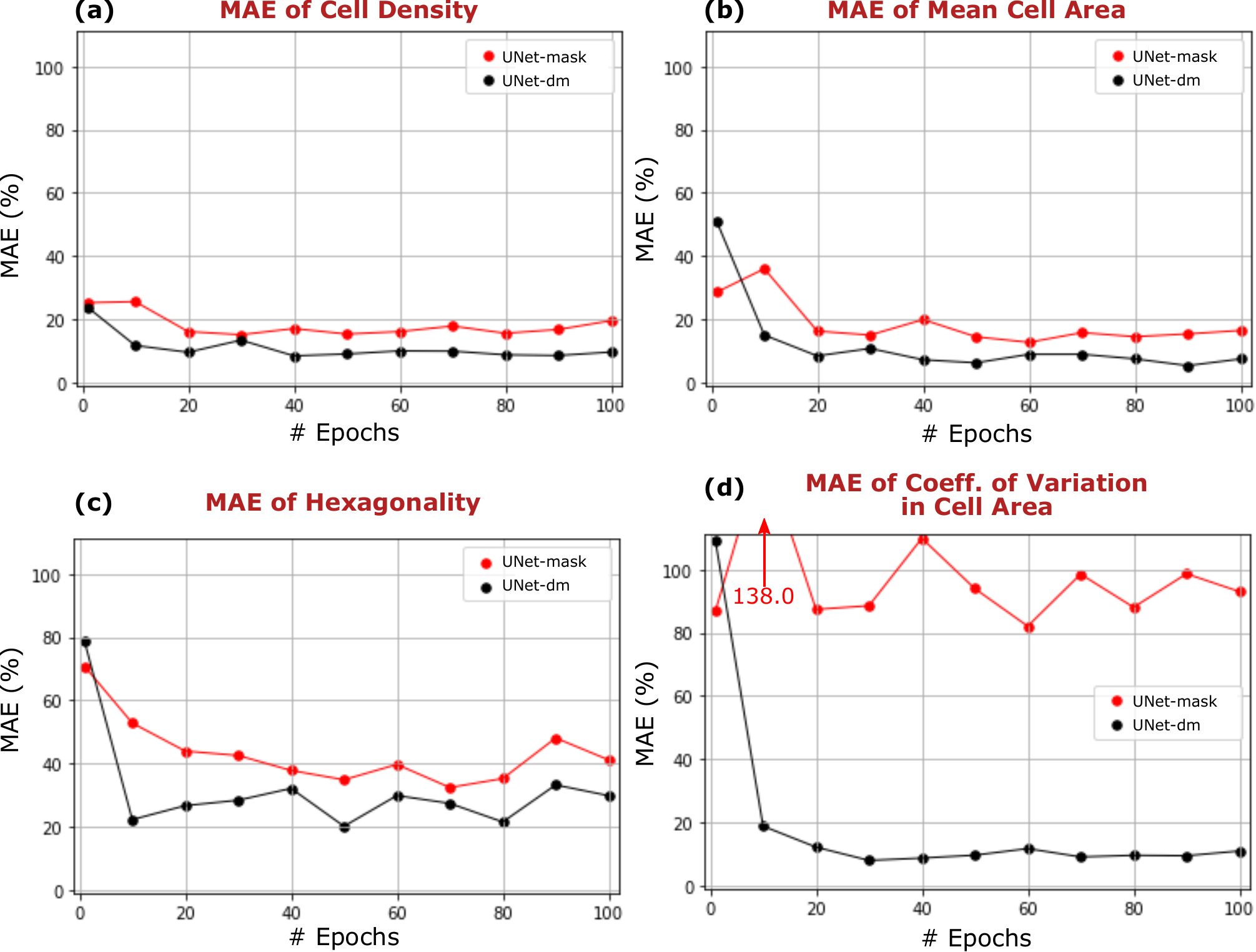}
 \caption{MAE calculated after each 10 training epochs of \textbf{(a)} Mean Cell Area, \textbf{(b)} Coefficient of Variation in Cell Area, \textbf{(c)} Cell Density and \textbf{(d)} Hexagonality. The graphs show the fast convergence and lower errors of the relevant morphometric parameters calculated using the UNet-dm compared to those calculated with the UNet-mask.}
 \label{fig:quantitative_comparison}
\end{figure}

The analysis of how the MAE of the morphometric parameters calculated on the testing set evolved every ten epochs of the training process shows a significant difference between the two versions of the UNet model. Fig.~\ref{fig:quantitative_comparison} reveals that the UNet-mask approach does not achieve the same performance as the UNet-dm even after 100 epochs. In sharp contrast, the UNet-dm method quickly converges to optimal performance with much lower MAEs for all parameters, but especially for the CV shown in Fig.~\ref{fig:quantitative_comparison}(d) in which the large error is due to many incorrectly segmented small cells.

% The analysis of how the MAE of the morphometric parameters calculated on the testing set evolved every ten epochs of the training process shows a significant difference between the two versions of the UNet model. On the one hand, Fig.~\ref{fig:quantitative_comparison} reveals that the MAE of each morphometric parameter decreases during the entire training stage for the UNet-mask despite not achieving the same. On the other hand, the UNet-dm, the MAE plots get flattened in the earliest epochs, with no significant improvement in the latest ones, which means that more epochs do not guarantee an improvement of the model. It demonstrates that the UNet-dm converges faster than the UNet-mask, reaching accurate results after only a few epochs. Also, the MAE of the UNet-dm model is considerably lower than the MAEs of the UNet-mask in each datapoint, giving the proposed method much better results.

\subsection{Performance evaluation 1: Qualitative analysis.}

\begin{figure}
\centering
 \includegraphics[width=0.8\linewidth]{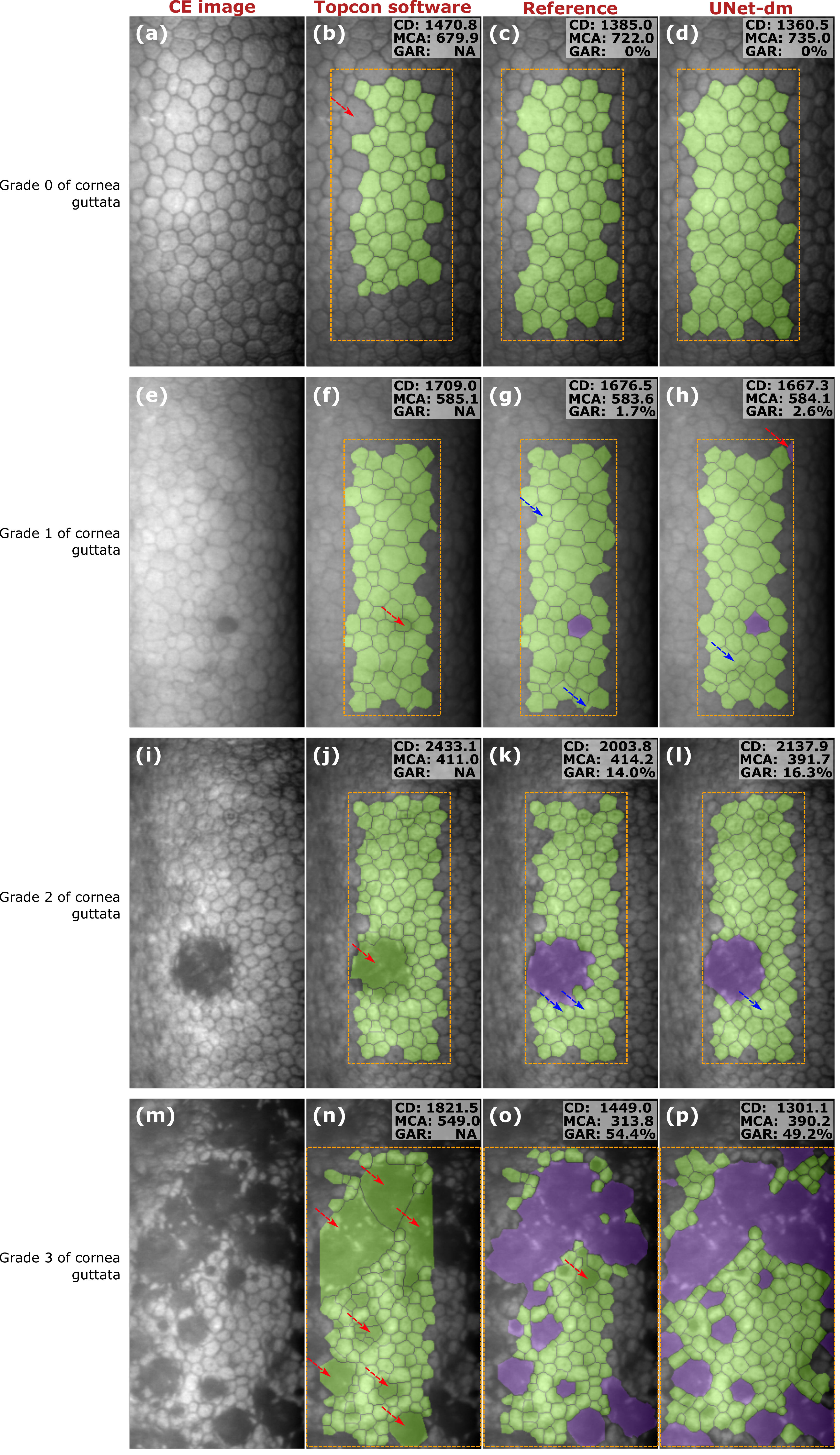}
 \caption{Qualitative analysis. From top to bottom: images from four different stages of cornea guttata. From left to right: the CE image, the Topcon software segmentation, the manual reference segmentation, and the predicted segmentation using the proposed method. Red arrows indicate examples of misclassified regions and blue arrows examples of inaccurately segmented regions. The orange bounding boxes indicate the analyzed ROI in each example.
}
 \label{fig:qualitative}
\end{figure}

Fig.~\ref{fig:qualitative} shows several representative results of four grades of endothelial dysfunction, demonstrating the robustness of the proposed method in different scenarios. The first column shows the specular microscopy images acquired with the Topcon SP-3000P specular microscope. The original segmentation performed by the microscope \textit{CellCount} software is shown in the second column. Column three contains their respective manually corrected segmentation. Finally, the automatic segmentations performed by the proposed method are shown in the fourth column. At the top-right corner of each segmentation, there are two of the main morphometric parameters: cell density (in cells/mm$^2$) and mean cell area (in $\mu \mathrm{m}^2$), in addition to guttae area ratio (in $\%$). It is noteworthy that the analyzed area is not the entire image but a ROI defined by the bounding box of the manually segmented area, which are surrounded by orange boxes in Fig.~\ref{fig:qualitative}.

The specular microscopy image of grade-0 cornea guttata (no guttae) in Fig.~\ref{fig:qualitative}(a) has higher quality and contrast than the other examples in the other rows. Here, the cell boundaries are easily detected. Therefore, the three segmentation results have similar performance. However, each method segmented slightly different areas, which produced minor discrepancies between the estimated morphometric parameters. For instance, the red arrow in Fig.~\ref{fig:qualitative}(b) points to a large cell not contained in the Topcon software segmentation. However, it was included with the annotation software in the reference and accurately segmented by the proposed UNet-dm model. %alongside several more cells. 

The second row of Fig.~\ref{fig:qualitative} shows a grade-1 cornea guttata image. In the Topcon segmentation, the red arrow indicates a small gutta, which the software erroneously classified as a cell. The reference segmentation includes it correctly as a gutta, as shown in Fig.~\ref{fig:qualitative}(g). However, even the reference segmentation may have errors from the manual annotation process. The blue arrows in the reference segmentation indicate two merged cells and an inaccurately segmented cell. The proposed method was able to correct these two issues (Fig.~\ref{fig:qualitative}(h)). However, the lack of uniform illumination in the periphery may lead to incorrect classification, such as the small dark region detected as a gutta indicated by the red arrow. The blue arrow points to a region that contains two merged cells due to the low contrast.

Grades 2 and 3 of corneal guttata are related to large guttae, like those pointed out with red arrows in the third and fourth row of Fig.~\ref{fig:qualitative}. The Topcon software erroneously detected them as large cells, which produce an overestimated MCA alongside any other morphometric parameter that depends on the cell area. In the reference segmentation of the severe case (Fig.~\ref{fig:qualitative}(o)), there is a medium-size gutta indicated with the red arrow that was not well classified. Nevertheless, the proposed UNet-dm method identified it correctly along with the larger guttae throughout the image and the cells.

\subsection{Performance evaluation 2: Quantitative results.}

\begin{figure}
\centering
 \includegraphics[width=0.8\linewidth]{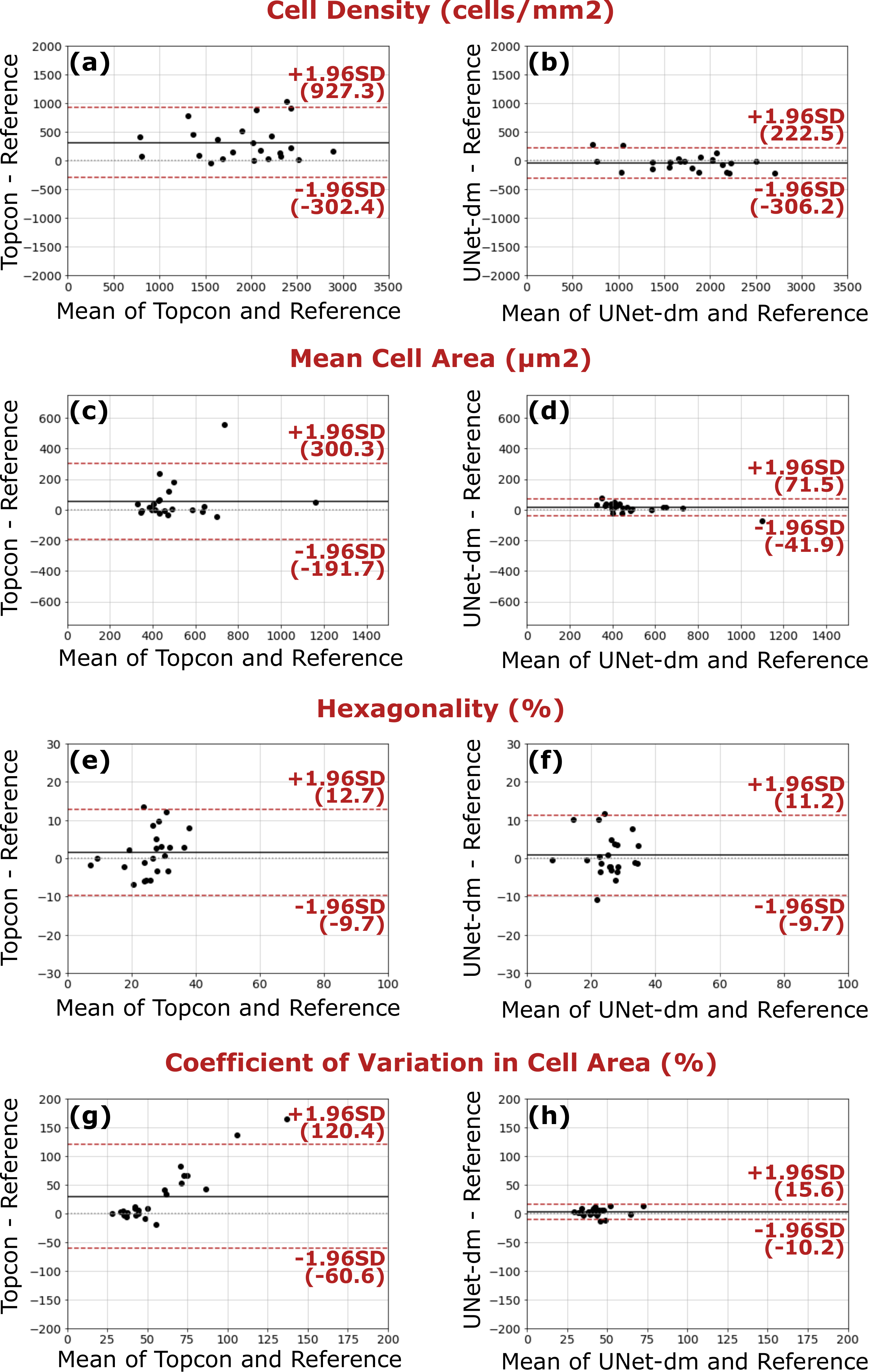}
 \caption{Bland-Altman plots of the main morphometric parameters of CE images. The first column shows the comparison between the original segmentation performed by the microscope and the manual corrected segmentation, i.e. the ground-truth reference, and the second column shows the comparison between the CNN-based proposed method and the ground-truth reference.}
 \label{fig:ba}
\end{figure}

We assume manually corrected segmentation as a ground-truth to compare with the results obtained from the proposed method. To assess the agreement between our proposed method and the reference, we used Bland-Altman plots (second column of Fig.~\ref{fig:ba}). More specifically, we evaluated the mean difference $\bar{d}$ between the two methods and determined the confidence interval CI = [$\bar{d}$ - 1.96 SD, $\bar{d}$ + 1.96 SD], within which we expect the 95\% of the differences to be contained. As is typical in a Bland-Altman plot, we assume the mean of the two measurements as the best estimation of the true value. The Bland-Altman plot displays the difference between the two methods against their mean. We carried out the same analysis to evaluate the agreement between the Topcon microscope software and the reference (first column of Fig.~\ref{fig:ba}).

Regarding the CD, the Bland-Altman plot of the Topcon segmentation and the reference shows a mean difference $\bar{d}$ = 312.4 cells/mm$^2$, which means the original segmentation heavily overestimates this parameter. This overestimation is due to dystrophic regions misclassified as cells, as shown with red arrows in the second column of Fig.~\ref{fig:qualitative}. The 95\% CI ranges from -302.4 to 927.3 cells/mm$^2$, which is considerably wide for a parameter crucial for assessing the CE. However, the bias between the proposed method and the reference is not zero but acceptable. On average, the proposed method measures -41.9 cells/mm$^2$ than the manual reference with 95\% CI [-306.2, 222.5] cells/mm$^2$. These results indicate good agreement between the proposed segmentation method and the manual reference with a relatively low CI, especially over an extensive CD range.

The analysis with the Bland-Altman plot of MCA between the Topcon software and the manual reference indicates an over-estimation by the Topcon software with a mean difference $\bar{d}$ = 54.3 $\mu m^2$, and 95\% CI in the range of [-191.7, 300.3] $\mu m^2$. This disagreement is due to large pathological regions classified as cells, like those being pointed out with red arrows in Fig.~\ref{fig:qualitative}(j,n). The results of comparing the proposed method and the reference yield a mean difference $\bar{d}$ =  14.8 $\mu m^2$, which is close to zero, and 1.96SD = $\pm$ 56.6 $\mu m^2$, indicating a good agreement between the two methods. 

As the HEX\% is not a cell size-dependent parameter, we expect it to be less affected by cornea guttata. The CIs of the two HEX\% Bland-Altman plots are quite similar. In comparison to the manual reference, the Topcon software produces 1.96SD = $\pm$ 11.2\% with $\bar{d}$ = 1.5\%, while the proposed method 1.96SD = $\pm$ 10.4\% and $\bar{d}$ = 0.8\%. They both confirm a good agreement with the ground truth.

The CV\% parameter yields a notable difference between the Bland-Altman plots. The mean difference between the Topcon segmentation and the reference was $\bar{d}$ = 29.9\%, with 95\% CI ranging from -60.6\% to 120.4\%. This plot seems to follow an upward trend; i.e., the error increases as the measured value is higher. On average, the Topcon segmentation tends to overestimate this parameter. Conversely, there is a good agreement level between the proposed method and the reference with a mean difference $\bar{d}$ = 2.7\% and a 95\% CI between -10.2\% and 15.6\%. 

Finally, the microscope often segments guttae as large cells or multiple small cells for dysfunctional regions. As a result, there is no information calculated with the Topcon software about these dystrophies. In the mild level of cornea guttata, the mean difference of GAR between the estimated values from the proposed method and the reference is $0.08 \pm 0.32$ \%. For the moderate level, the mean difference is $3.57 \pm 3.83$ \%, and for severe cases, it is $7.80 \pm 8.19$ \%. In all cases, these results indicate that the proposed method can reliably assess the percentage of area covered by guttae.

\section{Conclusions}

Automated corneal endothelium assessment in cornea guttata is challenging due to various factors, including adequate cell visibility, nonuniform illumination, and guttae appearance. Moreover, obtaining accurately defined segmentation boundaries from manual references for training is complicated. For this reason, we proposed a fast-converging UNet-based regression method trained to produce a signed distance map that requires minimal post-processing to calculate corneal endothelium morphometric parameters, reducing the complexity of the training process. Our results show that this method works sufficiently well, even with the relatively small dataset, and has the potential for improving the way specular microscopy images of the corneal endothelium are analyzed. Future work involves further validation on a larger dataset, exploring the impact of preprocessing strategies, like illumination correction, and the use of these methods in the clinical setting. The newly proposed guttae area ratio parameter may prove helpful for classifying Fuchs' dystrophy into different stages. 

\begin{backmatter}

\bmsection{Funding}
Ministerio de Ciencia, Tecnología e Innovación (763-2021), Universidad Tecnológica de Bolívar (CI2021P02), Agencia Estatal de Investigación del Gobierno de España (PID2020-114582RB-I00/ AEI / 10.13039/501100011033).

\bmsection{Acknowledgments}
This work has been partly funded by Ministerio de Ciencia, Tecnología e Innovación, Colombia, Project 124489786239 (Contract 763-2021), Universidad Tecnológica de Bolívar (UTB) Project CI2021P02, and Agencia Estatal de Investigación del Gobierno de España (PID2020-114582RB-I00/ AEI / 10.13039/501100011033). J. Sierra thanks UTB for a post-graduate scholarship. 

\bmsection{Disclosures} The authors declare no conflicts of interest.

\bmsection{Data availability} Data underlying the results presented in this paper are not publicly available at this time but may be obtained from the authors upon reasonable request.

\bmsection{Code availability} Code for the networks and their weights are available at~\cite{Sierra2022repo}.

\end{backmatter}

% \newpage
%%%%%%%%%%%%%%%%%%%%%%%%%%%%%%%%%%%%%%%%%%%%%%%%%%%%%
\bibliography{sample}

\end{document}